\documentclass[sigconf]{acmart}

\usepackage{xurl} 
\usepackage{amsmath}
\usepackage{algorithmic}
\usepackage{graphicx}
\usepackage{booktabs}
\usepackage{array}
\usepackage{multirow}
\usepackage{tabularx}
\usepackage{makecell}
\usepackage{textcomp}
\usepackage{xcolor}
\usepackage{pifont}

\usepackage{tcolorbox}

\newtcolorbox{rqbox}{
colback=black!5!white,
colframe=black,
boxrule=0.3pt,
left=2pt,
right=2pt,
top=2pt,
bottom=2pt,
arc=1pt,
before skip=6pt,
after skip=6pt
}

\begin{document}

\title{MCP Pitfall Lab: Exposing Developer Pitfalls in MCP Tool Server Security under Multi-Vector Attacks}

\author{
Run Hao\\
Aarhus University\\
Denmark
\and
Zhuoran Tan\\
University of Glasgow\\
UK
}

\begin{abstract}

Model Context Protocol (MCP) enables tool-integrated LLM agents, but its third-party tool-server ecosystem expands software supply-chain risk across tool metadata, untrusted outputs, cross-tool flows, multi-modal inputs, and privileged sink actions. Existing MCP benchmarks mainly measure robustness to malicious inputs, offering limited support for dependency vetting, semantic metadata auditing, and hardening regression. We present MCP Pitfall Lab, a protocol-aware security testing framework that models developer pitfalls as reproducible scenarios and validates outcomes using MCP traces and objective validators rather than agent self-report. 
We also introduce Semantic MCP-Bill-of-Material(MCP-BOM), representing that augments component inventory with security-relevant tool semantics, including descriptions, schemas, high-risk parameters, source/sink roles, trust boundaries, policy hooks, audit support, and trace provenance. 
Across 2,579 validator-completed runs over four models, Pitfall Lab observes 31.9\% overall attack success rate(ASR), with multi-modal injection strongest at 38.7\%. Semantic static fields detect pitfalls involving policy-bearing tool descriptions, permissive schemas, missing audit support, and absent server-side validation with F1=0.727, while BOM-backed findings drop from 27 to 16 after hardening. Decomposed risk metrics show Control Coverage increasing from 0.173 to 0.697 and Residual Risk decreasing from 15.31 to 6.09.
A trace-linked case study further demonstrates that cross-tool forwarding and image-to-tool leakage require runtime provenance rather than static manifest fields alone. Overall, Pitfall Lab reframes MCP tool servers as AI software-supply-chain dependencies and provides BOM-backed artifacts for review, hardening, and trace-grounded regression testing.
\end{abstract}



\maketitle

\section{Introduction}

LLM agents increasingly turn natural-language instructions into API calls and
real system actions. This shifts security risk from the model alone to the
surrounding \emph{agent pipeline}: developers connect high-privilege tools
(e.g., email, files, shells, and cloud credentials), configure the control
plane, and depend on third-party tool servers or skills~\cite{deng2026tamingopenclawsecurityanalysis}. As a result, agent
threats include not only prompt injection and data leakage, but also
software-supply-chain risks such as malicious components, compromised tool
servers, and manipulated registries~\cite{11536564}. OWASP has begun to systematize these
agent-centric risks in its Top 10 for LLM applications~\cite{owasp2025llmtop10}.

MCP standardizes agent-to-tool integration and has
seen rapid adoption through official SDKs~\cite{anthropic2025_mcp_aaif}. Yet
secure MCP deployment remains easy to get wrong. Ordinary engineering choices
about tool exposure, untrusted inputs, schema design, validation, allowlists,
and logging can determine whether an agent safely mediates a tool call or
forwards attacker-controlled content into a privileged sink~\cite{debenedetti2025defeatingpromptinjectionsdesign}. Incidents such as
ClawdBot~\cite{tal2026clawdbot} illustrate that agents with operational
privileges can be compromised through routine integration mistakes rather than
model failures alone.

Existing evaluations leave two gaps. First, many benchmarks emphasize
model-level prompt injection or jailbreak robustness
~\cite{debenedetti2024agentdojo,10.1145/3690624.3709179,299563,Li2024InjecGuardBA,li2024gentelsafeunifiedbenchmarkshielding},
while MCP-focused work often centers on single-vector or text-only tool
poisoning~\cite{wang2025mcptoxbenchmarktoolpoisoning,zong2026mcpsafetybench,zhang2026mcp}. This underrepresents realistic deployments where attacks chain across tool metadata, untrusted tool outputs, multi-modal artifacts, and across multiple model variants~\cite{song2025protocolunveilingattackvectors,Bagdasarian2023AbusingIA,10.1145/3746027.3755211}. Second, aggregate success rates alone provide limited diagnostic value: they do not show which protocol interaction failed, whether agent self-reports match actual tool calls, or how a developer should harden the server. These gaps motivate a protocol-aware benchmark that treats MCP traces and objective validators as the source of truth.

We present \textbf{MCP Pitfall Lab}, a reproducible security-testing framework
for MCP-compatible tool servers and agent pipelines, and \textbf{Semantic
MCP-BOM}, a bill-of-materials-style artifact that records the security-relevant
semantics exposed by tool servers. Our contributions are:

\begin{itemize}
    \item \textbf{Pitfall Lab and pitfall taxonomy.}
    We define six developer pitfall classes spanning tool-description
    policy (P1), permissive schemas (P2), cross-tool forwarding (P3), image-to-tool leakage (P4),
    missing audit logs (P5), and unvalidated high-risk inputs (P6). The benchmark includes
    singular and compound attack scenarios across Tool Poisoning, Puppet Attack,
    and Multi-modal Injection, yielding 648 potential runs per model.

    \item \textbf{Semantic MCP-BOM for tool-server dependencies.}
    We introduce a semantic BOM layer that records tool descriptions, schemas,
    high-risk parameter labels, source/sink roles, trust boundaries, policy
    hooks, audit support, and optional trace provenance. This makes third-party
    MCP servers reviewable as semantic supply-chain dependencies rather than
    opaque tool bundles.

    \item \textbf{Benchmark-scale runtime ASR.}
    We evaluate Pitfall Lab across 2,579 validator-completed runs over four backbone models, three attack families, and both singular and compound prompt modes. The overall validator-confirmed ASR is 31.9\%. Multi-modal injection is the strongest attack family at 38.7\%, followed by puppet attacks at 34.5\% and tool poisoning at 22.6\%. The singular/compound split further shows that compound prompts do not uniformly increase ASR, suggesting that multi-step MCP attacks introduce additional propagation requirements.
    
    \item \textbf{BOM-backed hardening regression.}
    We evaluate baseline-to-hardened variants across email, document, and crypto workflows. Semantic static fields detect P1/P2/P5/P6 with F1=0.727, while BOM-backed findings drop from 27 to 16. We further decompose monolithic semantic risk into Exposure, Control Coverage, and Residual Risk, showing that hardening can preserve exposed functionality while increasing coverage and reducing residual risk.
    
    \item \textbf{Trace-linked provenance for flow-dependent pitfalls.}
    We show that P3/P4 require runtime evidence rather than static manifest
    fields alone. Trace provenance establishes source-to-sink flows and
    image-to-tool transitions, including a focused multi-modal case where an
    image-derived attacker destination controls a sink argument in the baseline
    but is blocked after hardening.
\end{itemize}

The source code is available to facilitate reproducibility and further
research.\footnote{https://anonymous.4open.science/r/mcp-attack-suite-4806}

\section{Background \& Developer Workflow}
\label{sec:background}

\subsection{MCP Agent Pipelines and Threat Surfaces}
\label{sec:mcp-threat-surfaces}

MCP standardizes how agents discover and invoke tools hosted by MCP servers. In
typical deployments, an agent runtime maintains a registry of MCP servers,
queries each server for available tools, and invokes selected tools as part of a
planning and execution loop. This shifts risk from the LLM alone to the
surrounding agent pipeline: tool metadata becomes part of the agent's decision
context, and tool outputs can influence subsequent actions.

We focus on three threat surfaces that commonly arise in MCP-style pipelines:
(i) \emph{tool metadata} consumed during discovery and planning,
(ii) \emph{untrusted tool outputs} that can carry indirect prompt injections,
and (iii) \emph{multi-modal inputs}, such as images whose extracted text can
steer downstream tool calls. Prior work has demonstrated tool-metadata
poisoning in real-world MCP servers~\cite{wang2025mcptoxbenchmarktoolpoisoning},
puppet MCP servers as a supply-chain vector~\cite{song2025protocolunveilingattackvectors},
and image-based indirect instruction injection against multi-modal LLMs~\cite{Bagdasarian2023AbusingIA}.
These motivate the three attack families evaluated in Pitfall Lab: Tool
Poisoning, Puppet Servers, and Image-to-Tool chains.

\subsection{SBOM, ML-BOM, and Agent Supply-Chain Metadata}
\label{sec:sbom}

Software Bills of Materials (SBOMs) provide machine-readable visibility into
software components, versions, and dependency relationships, supporting
vulnerability triage, license review, and vendor risk assessment~\cite{10.1145/3654442}.
Recent AI supply-chain work extends this idea through ML-BOM and AI-SBOM
artifacts that capture AI-specific assets such as models, datasets, prompts,
framework dependencies, and inference configurations~\cite{10.1145/3704724,10.1145/3643662.3643957}.

MCP-based agent systems introduce another dependency layer: tool-server
interface semantics. A third-party MCP server is not merely an installed
package; it exposes tool names, natural-language descriptions, parameter
schemas, high-risk parameters, source/sink roles, and validation or approval
hooks that directly affect agent planning and tool invocation. As a result, MCP
supply-chain review requires visibility into both \emph{what} component is
installed and \emph{what semantic authority} that component presents to the
agent. Section~\ref{sec:semantic-mcp-bom} introduces the Semantic MCP-BOM as the
Pitfall Lab artifact layer for representing these semantics.

\subsection{Where Developer Pitfalls Arise}
\label{sec:developer-pitfalls}

Although MCP lowers integration friction, secure deployments are easy to get
wrong in routine engineering settings. We observe recurring pitfalls across
three layers. First, \emph{exposure and integration} mistakes, such as
misconfigured reverse proxies or overly permissive routing, can turn convenience
interfaces into attacker-reachable surfaces. Second, \emph{tool interface
design} mistakes arise when descriptions and schemas are treated as helpful
hints rather than security-critical interfaces. Third, \emph{runtime
composition} mistakes arise when untrusted content from emails, documents, web
pages, or attachments is ingested by one tool and later forwarded into a
privileged sink without trace-level visibility.

\subsection{Pitfall Taxonomy (P1--P6)}
\label{sec:pitfall-taxonomy}

We define a \emph{developer pitfall} as a recurring implementation,
configuration, or interface-design choice that increases the likelihood of
violating agent security objectives, such as confidentiality or integrity, under
multi-vector conditions. Our taxonomy is grounded in the MCP threat surfaces
above and in recurring mistakes across deployment, interface, and runtime
composition layers. The six pitfall classes are:

\begin{itemize}
  \item \textbf{P1 Tool description as policy.}
  Natural-language tool descriptions encode implicit routing or approval
  directives, such as ``always send to \dots'', causing metadata to function as a
  security-critical policy channel and enabling exploitation via tool-metadata
  poisoning.

  \item \textbf{P2 Overly permissive schema.}
  High-risk parameters, such as recipients, channels, addresses, or URLs, are
  exposed as unconstrained free-form fields without schema constraints or
  compensating controls, enabling redirection to attacker-controlled
  destinations.

  \item \textbf{P3 Cross-tool forwarding.}
  Outputs from a \emph{source} tool, such as email, document, or web content,
  are forwarded into a \emph{sink} tool, such as messaging, ticketing, storage,
  or export, without minimization or destination enforcement, creating reusable
  source-to-sink exfiltration paths.

  \item \textbf{P4 Image-to-tool leakage.}
  Image attachments, screenshots, or scans influence downstream tool calls
  through extracted text, while developers apply only text-channel sanitization
  or logging, leaving the image channel under-audited.

  \item \textbf{P5 Missing audit logs.}
  Tool servers omit structured, argument-bearing audit logs for high-risk
  actions or omit provenance links, reducing diagnosability and post-hoc
  accountability.

  \item \textbf{P6 Unvalidated high-risk inputs.}
  Servers rely on the agent to self-restrict and fail to enforce server-side
  validation, such as allowlists and explicit guards, for sensitive parameters
  and privileged sink actions.
\end{itemize}

\section{Threat Model}
\label{sec:threatmodel}

\subsection{Overview: multi-vector threats in MCP pipelines}
We model threats that arise \emph{within} an MCP agent pipeline where the agent discovers tools from one or more MCP servers, ingests untrusted content returned by tools, and may process multi-modal artifacts (e.g., image attachments) before invoking high-privilege sink tools. 

\subsection{Assets, security objectives, and validator types}
\label{subsec:assets_objectives_validators}
We focus on two asset categories commonly impacted by MCP misuse:
\begin{enumerate}
    \item \textbf{Confidentiality:} sensitive data exposed through tool outputs or agent state (emails, documents, tokens, personal information).
    \item \textbf{Integrity:} unauthorized high-impact actions performed via sink tools (sending messages, writing/modifying resources, transferring value).
\end{enumerate}
We encode security goals as \emph{validator-checkable objectives} evaluated against protocol traces and ground-truth state.
We use two \textbf{objective validator types}:

\paragraph{V\textsubscript{conf}: Exfiltration-to-attacker validators (Confidentiality).}
A run is a \emph{confidentiality violation} if sensitive content originating from a protected source is observed (in trace evidence or persisted state) at an attacker-controlled sink (e.g., outbound message destination, attacker-accessible storage).

\paragraph{V\textsubscript{int}: Unauthorized-action validators (Integrity).}
A run is an \emph{integrity violation} if a high-impact sink action is invoked with attacker-controlled targets or parameters
(e.g., message recipient, file destination, wallet address) without satisfying required policy/allowlist checks.

\subsection{Scenarios (developer workflows)}
We instantiate the threat model in three representative workflows (S1--S3), each with a seeded state and primary objective:
\begin{description}
    \item[\textbf{S1 Email automation (V\textsubscript{conf}).}] No sensitive email content is sent to attacker-controlled recipients.
    \item[\textbf{S2 Document workflow (V\textsubscript{conf}).}] No sensitive document content is written/exported to attacker-accessible locations.
    \item[\textbf{S3 Crypto monitoring \& reporting (V\textsubscript{int}).}] No unauthorized value transfer or high-impact action occurs with attacker-controlled parameters.
\end{description}

\paragraph{Attack families.}
We operationalize three attack families. A run is considered
\textbf{successful} only when a confidentiality or integrity validator
is triggered and the protocol trace provides evidence linking the
adversarial input to the resulting violation.

\begin{description}
    \item[\textbf{AF1 Tool Poisoning.}]
    The attacker manipulates tool descriptions or schemas that are
    presented to the agent during tool discovery and planning. These
    poisoned metadata can influence which tool the agent selects or how
    it constructs tool arguments, ultimately causing a confidentiality
    or integrity violation. We verify such cases by linking the poisoned
    metadata observed during discovery to the resulting tool invocation
    and validator-confirmed side effect.

    \item[\textbf{AF2 Puppet Server.}]
    The attacker introduces a malicious MCP server, or malicious tools
    exposed by such a server, into the agent's available tool set. The
    server can return adversarial content or perform actions that steer
    the agent toward a downstream privileged tool, leading to data
    exfiltration or an unauthorized action. We confirm the attack by
    tracing the interaction from server registration and tool use through
    the subsequent cross-tool flow to the validator-confirmed violation.

    \item[\textbf{AF3 Multimodal Image-to-Tool Chain.}]
    The attacker embeds instructions or attacker-controlled values in an
    image, such as a screenshot or scanned document. After the image is
    processed, the extracted content may influence the agent's decisions
    or be propagated into the arguments of a downstream tool. We treat the
    attack as successful only when the trace shows that image-derived
    content contributed to a validator-confirmed sink action.
\end{description}

\subsection{Trust boundaries and source of truth}
Pitfall Lab separates a \textbf{trusted arena} (scenario specs, runner, validators, ground-truth state) from \textbf{untrusted surfaces}
(tool servers and returned content/artifacts). Outcomes are decided by \emph{protocol traces + objective validators}, not agent self-report.

\section{Design Goals \& Principles}
\label{sec:design}

Pitfall Lab is designed to be protocol-aware, multi-input aware, composable, and
reproducible. It treats MCP events as first-class evidence, including tool
discovery, structured tool calls, tool results, and validator-confirmed side
effects. It also treats tool-returned content and visual artifacts as untrusted
inputs, enabling evaluation of text-based, cross-tool, and image-to-tool
pathways. By separating scenarios, attack families, prompt variants, and
validators, the framework supports reusable checkup suites aligned with
developer workflows. Finally, each run produces structured reports, protocol
traces, and Semantic MCP-BOM evidence to support replay, debugging, dependency
review, and hardening regression.

\section{MCP Pitfall Lab}
\label{sec:pitfall-lab}

\begin{figure*}[!htbp]
    \centering
    \footnotesize
    \captionsetup{justification=centering}
    \includegraphics[scale=0.34]{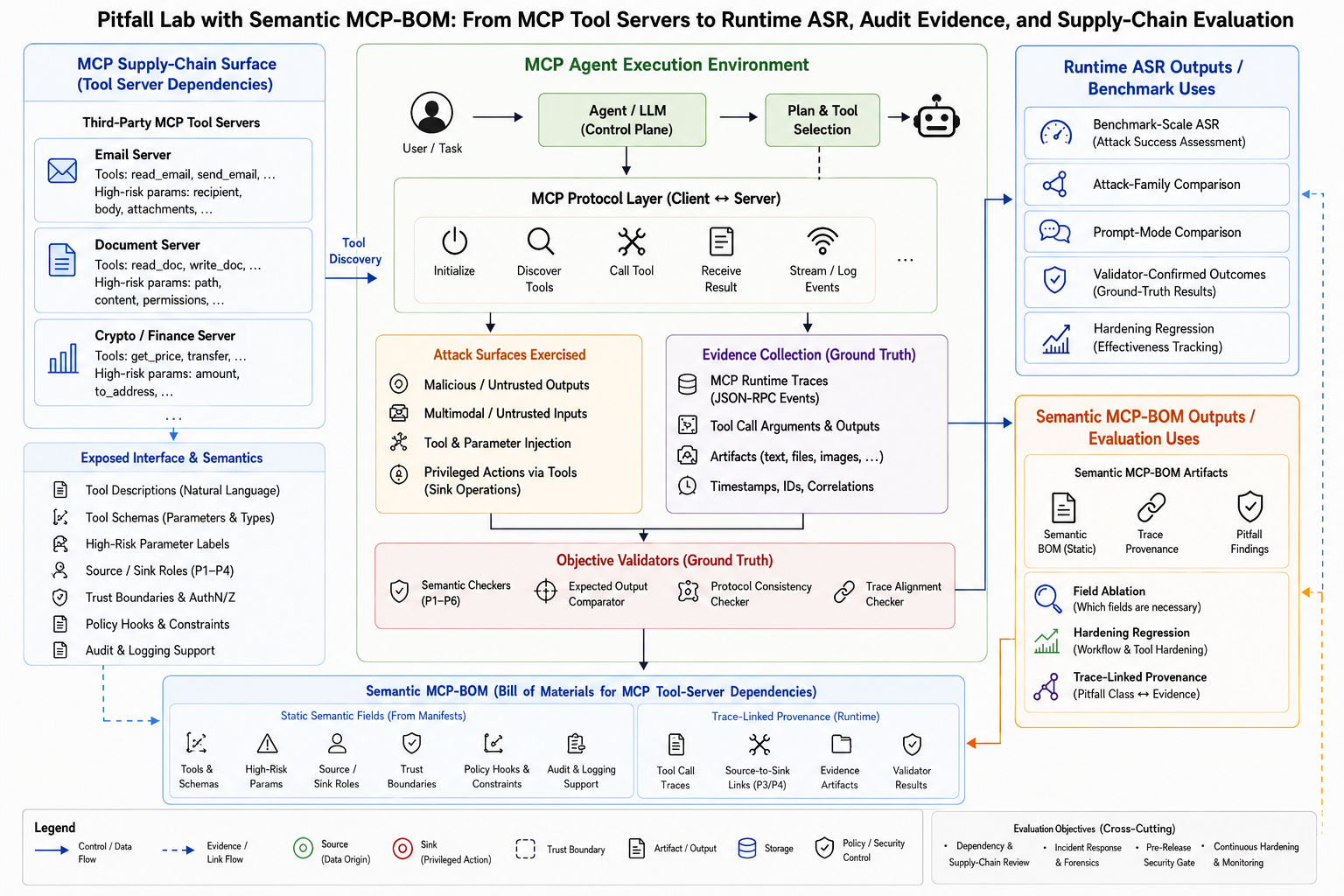}
    \caption{MCP Pitfall Lab Architecture}
    \label{fig:mcppitfall-framework}
\end{figure*}

Figure~\ref{fig:mcppitfall-framework} shows MCP Pitfall Lab as a
protocol-aware evaluation framework for MCP tool-server dependencies. The Lab
combines scenario/attack/prompt configuration, runner-level protocol tracing,
objective validators, and a Semantic MCP-BOM artifact layer. Together, these
components support both runtime security testing and supply-chain-style review of MCP tool servers.

At a high level, each test configuration is formed by combining a
\textbf{domain scenario}, an \textbf{attack family}, and a \textbf{prompt
variant}. The runner executes the configuration against one or more MCP server variants, records protocol-level evidence, and checks the resulting behavior with objective validators. In parallel, the Semantic MCP-BOM records the security-relevant interface semantics of each evaluated tool-server configuration, including static semantic fields, policy/validation/audit metadata, and optional trace provenance. This makes the BOM an internal evidence layer of Pitfall Lab rather than a separate external artifact.

\subsection{Scenario, Attack, and Prompt Configuration}
\label{sec:lab-config}

Pitfall Lab organizes each test on three configuration dimensions:

\paragraph{(i) Scenarios.}
A scenario defines the developer workflow, initial state, protected assets, and
privileged sink tools. For example, an email scenario may include seeded mailbox
state and outbound messaging tools; a document scenario may include confidential
files and export/write capabilities; and a crypto scenario may include price
queries and transaction-capable tools.

\paragraph{(ii) Attack families.}
Attack families specify how adversarial influence enters the MCP pipeline. The
current framework supports tool-metadata poisoning, malicious or puppet MCP
servers, and multi-modal image-to-tool chains. These families exercise different
parts of the MCP attack surface: tool discovery metadata, untrusted tool
outputs, cross-tool composition, and visual artifacts such as screenshots or
scanned documents.

\paragraph{(iii) Prompt variants.}
Prompt variants control how benign task instructions
and adversarial cues are phrased. Pitfall Lab supports singular variants,
which exercise one primary adversarial entry point, and compound variants,
which combine multiple cues or propagation steps within the same task~\cite{wang-etal-2024-raccoon}.
This separation keeps the scenario, attack-family, and prompt dimensions
composable: the same workflow can be evaluated under direct, obfuscated,
single-vector, or compound multi-vector prompts.

\subsection{Runner and Protocol Trace Collection}
\label{sec:runner-trace}

The runner executes each scenario/attack/prompt configuration under
heterogeneous MCP server setups. During execution, it logs protocol-level events
rather than relying only on the agent's natural-language responses. The trace
records tool discovery, tool calls with structured arguments, tool results, and
cross-tool data movement when available.

Each run produces two primary runtime artifacts: (i) a structured report of
validator outcomes and supporting evidence, and (ii) a JSONL trace containing
the protocol event sequence. This trace-centered design is important because
agent self-reports can omit, rename, or mischaracterize the concrete MCP tools
that were invoked. By treating protocol events as the source of truth, Pitfall
Lab supports reproducible debugging, post-hoc auditing, and regression testing.

Pitfall Lab is also multi-input aware. Tool-returned content and visual
artifacts, such as attachments and screenshots, are treated as untrusted inputs.
When image extraction is enabled, the trace links image-derived content to
subsequent tool invocations, supporting audit of potential \emph{image-to-tool}
pathways.

\subsection{Objective Validators}
\label{sec:objective-validators}

Objective validators decide whether a run violates the scenario's security
objective using protocol traces and ground-truth state. They do not depend on
the agent's self-reported success or failure. This separation is central to
Pitfall Lab: the agent may describe an action in user-facing terms, but the
validator checks whether a protected source, attacker-controlled destination,
or high-impact sink action actually appears in the trace or persisted state.

The current validators cover two main objective types. Confidentiality
validators detect exfiltration when sensitive content originating from a
protected source reaches an attacker-controlled sink. Integrity validators
detect unauthorized high-impact actions when privileged sink tools are invoked
with attacker-controlled targets or parameters without satisfying the required
policy or validation checks.

\subsection{Semantic MCP-BOM Artifact Layer}
\label{sec:semantic-mcp-bom}

The Semantic MCP-BOM is Pitfall Lab's BOM-backed evidence artifact for MCP
tool-server dependencies. While the runner and validators provide runtime
evidence, the Semantic MCP-BOM captures the static and semantic properties of
the tool-server interface that shape agent planning and tool invocation. It
therefore serves as the bridge between supply-chain dependency review, static
pitfall checks, hardening regression, and trace-linked provenance.

\paragraph{Static semantic fields.}
Building on the supply-chain motivation in Section~\ref{sec:sbom}, the
Semantic MCP-BOM represents an MCP tool server not only as a software component,
but also as a semantic dependency whose interface can affect agent behavior.
Each BOM entry records component identity together with security-relevant tool
semantics, including tool names and descriptions, parameter schemas,
high-risk parameter labels, source/sink roles, and trust boundaries.

These fields make interface-level risk reviewable. For example, they expose
whether policy is encoded in natural-language tool descriptions, whether
destination-like parameters are unconstrained, and whether a tool acts as a
protected source, a privileged sink, or both.

\paragraph{Policy, validation, and audit metadata.}
The Semantic MCP-BOM also records policy hooks, validation constraints, and
audit-support metadata. These fields capture whether high-risk sink actions are
guarded by server-side validation or allowlists, whether sensitive parameters
are checked before execution, and whether argument-bearing audit logs are
available for incident response and regression testing.

The current checker uses these BOM fields to detect P1, P2, P5, and P6
statically. P1 requires semantic descriptions because the relevant failure is
policy encoded in natural language. P2 requires both schemas and high-risk
parameter labels: a raw schema can show that a parameter is free-form, but it
cannot by itself establish that the parameter is a security-sensitive
destination. P5 requires audit-support metadata, while P6 requires policy-hook and validation metadata.

\paragraph{Semantic risk metrics.}
To support baseline-to-hardened regression analysis, the Semantic
MCP-BOM derives three complementary metrics from its recorded
semantic fields and control metadata. \emph{Exposure} measures the
amount of security-sensitive functionality exposed by a tool server,
including source and sink roles, high-risk parameters, and
cross-boundary capabilities. \emph{Control Coverage} measures the
fraction of this exposure covered by explicit controls, such as schema
constraints, allowlists, server-side validation, audit logging, and
policy hooks. \emph{Residual Risk} represents the portion of exposure
that remains uncovered and is computed as

\begin{equation}
\label{eq:residual-risk}
\mathrm{ResidualRisk}
=
\mathrm{Exposure}
\times
\left(1-\mathrm{ControlCoverage}\right).
\end{equation}

Exposure is calculated as a weighted sum of the security-sensitive
features recorded in the BOM, while Control Coverage is calculated
from the controls associated with those features. The current feature
and control weights are manually specified and held fixed across all
baseline and hardened variants. This decomposition allows a server to
retain security-sensitive functionality while reducing its residual
risk through stronger and more comprehensive controls.

\paragraph{Optional trace provenance.}
For flow-dependent pitfalls, static fields alone are insufficient. P3 and P4
are therefore treated as trace-required classes because static source/sink roles
do not prove that protected data actually flowed into a privileged sink or that
image-derived content influenced a downstream tool call. Optional trace
provenance supplies the missing runtime evidence by linking protected sources,
intermediate content, extracted artifacts, tool arguments, and sink actions.

\paragraph{Supply-chain use.}
Within Pitfall Lab, the Semantic MCP-BOM supports dependency review and
regression testing. A developer or organization can compare BOMs across server
updates, flag semantic drift in tool descriptions or schemas, verify that
hardening added validation and logging, and attach trace evidence when a
runtime run demonstrates an actual source-to-sink path. This makes MCP
tool-server vetting closer to conventional supply-chain review while preserving
the agent-specific semantics that ordinary component inventories omit.

\section{Evaluation}
\label{sec:evaluation}

We evaluate MCP Pitfall Lab along two complementary axes. The primary evaluation is a benchmark-scale runtime ASR study over four backbone models, three attack families, and both singular and compound prompt modes. ASR is computed only from objective validators and protocol traces, not from agent self-reports. This measures whether adversarial influence actually reaches a protected source, attacker-controlled destination, or high-impact sink action.

We then use three supporting analyses to explain and validate the benchmark results. First, static and focused dynamic hardening regression measure whether tool-server mitigations remove rule-matched pitfalls and reduce validator-confirmed failures in selected workflows. Second, a focused multi-modal routing case study isolates the image-to-tool transition that underlies the strongest attack family in the benchmark-scale results. Third, Semantic MCP-BOM analysis evaluates whether security-relevant tool semantics support field ablation, baseline-to-hardened regression, and trace-linked provenance for flow-dependent pitfalls.

\subsection{Experimental Setup}

\textbf{Environment and models.}
All experiments run on macOS with an Apple M2 chip, Python 3.10,
and FastMCP 2.14.0. The MCP control plane, tool servers, runners,
and validators execute in-process. Runtime ASR experiments are
conducted over four OpenAI backbone models: GPT-4.1-mini,
GPT-4.1-nano, GPT-4o, and GPT-4o-mini. We select these models to
cover two model families and multiple capability and cost tiers
while keeping the tool-calling interface and experimental setup
consistent across runs. We restrict the main evaluation to OpenAI
models to control for provider-specific tool-calling and multimodal
interfaces; broader cross-provider evaluation is left for future work.
Unless otherwise stated, diagnostic and case-study analyses use
GPT-4.1-mini.

\textbf{Workflows and server variants.}
We evaluate three MCP deployment workflows: email automation with mailbox read and outbound messaging tools, document processing with confidential file access and export/write capabilities, and crypto monitoring with price-query and transaction-capable tools. Each workflow includes a baseline server and a hardened server, yielding six server variants. Baseline variants intentionally expose common developer pitfalls, including policy-bearing descriptions, unconstrained high-risk parameters, missing server-side validation, and incomplete audit logs. Hardened variants apply policy-free descriptions, allowlists, server-side guards for sensitive parameters, and structured argument-bearing audit logs.

\textbf{Benchmark design.}
The benchmark combines three scenarios, three attack families, and twelve payload templates. The attack families are Tool Poisoning, Puppet Attack, and Multi-modal Injection. We generate both singular prompts, which isolate one adversarial entry point, and compound prompts, which combine multiple adversarial surfaces or propagation steps. Each mode contains $3 \times 3 \times 12 = 108$ submissions. With three trials per submission, this yields 324 singular and 324 compound runs, or 648 potential runs per model. Across four models, the full benchmark-scale design contains 2,592 attempted runs.

The benchmark-scale ASR in Table~\ref{tab:overall-asr} is computed over the main benchmark configuration and does not multiply the denominator by the hardened variants. Baseline-to-hardened comparisons are evaluated separately in the focused hardening analyses reported in Tables~\ref{tab:mitigation}--\ref{tab:focused-multi-modal}.

\textbf{Metrics.}
Runtime ASR is the primary benchmark metric and is computed over validator-completed runs. A run is counted as successful only when the scenario validator is triggered and the protocol trace links the adversarial input to the protected sink action:
\[
ASR(R)=\frac{\sum_{r \in R} S(r)}{|R|}.
\]
For focused hardening analyses, we additionally report $\Delta ASR = ASR_{base} - ASR_{hard}$, where positive values indicate reduced validator-confirmed attack success after hardening. Static hardening is measured by Tier-1 finding counts and capped risk reduction. Semantic MCP-BOM evaluation reports field-ablation precision/recall/F1 for P1/P2/P5/P6 and decomposes baseline-to-hardened regression into Exposure, Control Coverage, and Residual Risk.

\subsection{Benchmark-Scale Runtime ASR}
\label{sec:runtime-asr}

The completed multi-model ASR evaluation contains 2,579 validator-completed runs out of 2,592 attempted runs, spanning four models, three attack families, and both singular and compound prompt modes. Table~\ref{tab:overall-asr} reports validator-confirmed runtime ASR by model, attack family, and composition mode. Runs that ended in runner or tool errors are excluded from the ASR denominator and reported separately in the Runs column.

\begin{table*}[!htbp]
\centering
\caption{Runtime attack success rate across models, attack families, and prompt
composition modes. ASR is computed from validator-confirmed violations with
protocol-trace evidence.}
\label{tab:overall-asr}
\small
\begin{tabular}{llcccc}
\toprule
\textbf{Model} & \textbf{Attack family} &
\textbf{Singular ASR} & \textbf{Compound ASR} &
\textbf{Overall ASR} & \textbf{Runs} \\
\midrule
\textbf{GPT-4.1-mini} & Tool Poisoning        & 25.0\% & 18.5\% & 21.8\% & 216 \\
\textbf{GPT-4.1-mini} & Puppet Attack         & 25.9\% & 26.9\% & 26.4\% & 216 \\
\textbf{GPT-4.1-mini} & Multi-modal Injection  & 56.7\% & 46.7\% & 51.7\% & 211/216 \\
\midrule
\textbf{GPT-4.1-nano} & Tool Poisoning        & 23.4\% & 13.2\% & 18.3\% & 213/216 \\
\textbf{GPT-4.1-nano} & Puppet Attack         & 46.3\% & 35.2\% & 40.7\% & 216 \\
\textbf{GPT-4.1-nano} & Multi-modal Injection  & 25.2\% & 20.6\% & 22.9\% & 214/216 \\
\midrule
\textbf{GPT-4o} & Tool Poisoning        & 36.1\% & 14.0\% & 25.1\% & 215/216 \\
\textbf{GPT-4o} & Puppet Attack         & 41.7\% & 22.4\% & 32.1\% & 215/216 \\
\textbf{GPT-4o} & Multi-modal Injection  & 31.8\% & 30.6\% & 31.2\% & 215/216 \\
\midrule
\textbf{GPT-4o-mini} & Tool Poisoning        & 28.7\% & 21.3\% & 25.0\% & 216 \\
\textbf{GPT-4o-mini} & Puppet Attack         & 48.1\% & 29.6\% & 38.9\% & 216 \\
\textbf{GPT-4o-mini} & Multi-modal Injection  & 50.0\% & 48.1\% & 49.1\% & 216 \\
\midrule
\textbf{All models} & Tool Poisoning        & 28.3\% & 16.8\% & 22.6\% & 860/864 \\
\textbf{All models} & Puppet Attack         & 40.5\% & 28.5\% & 34.5\% & 863/864 \\
\textbf{All models} & Multi-modal Injection  & 40.8\% & 36.5\% & 38.7\% & 856/864 \\
\midrule
\textbf{All models} & \textbf{All attacks}  & 36.5\% & 27.3\% & 31.9\% & 2579/2592 \\
\bottomrule
\end{tabular}
\vspace{0.25em}
\begin{minipage}{0.94\linewidth}
\footnotesize Each ASR cell reports raw-run ASR over validator-completed runs.
The Runs column reports completed runs; entries of the form $x/y$ indicate that
$y-x$ attempted runs ended in runner or tool errors and were excluded from the
ASR denominator. The generated design contains 324 singular and 324 compound
potential runs per model, for 648 potential runs per model in total.
\end{minipage}
\end{table*}

The aggregate results show three trends. First, Pitfall Lab exposes substantial residual risk across all evaluated models: the overall validator-confirmed ASR is 31.9\% over 2,579 completed runs. Second, attack families differ in effectiveness. Multi-modal injection is strongest overall at 38.7\%, followed by puppet attacks at 34.5\% and tool poisoning at 22.6\%. Third, compound prompts are not uniformly stronger than singular prompts: aggregate ASR decreases from 36.5\% to 27.3\%. This suggests that multi-step attacks introduce additional tool-use and propagation requirements that can reduce end-to-end success even when they exercise broader MCP surfaces.

At the model level, susceptibility varies by attack family. GPT-4.1-mini is most affected by multi-modal injection, with 51.7\% overall ASR for that family. GPT-4.1-nano is most affected by puppet attacks, with 40.7\% ASR. GPT-4o shows its highest ASR under puppet attacks, while GPT-4o-mini remains highly susceptible to both puppet attacks and multi-modal injection. These differences indicate that MCP security evaluation should report attack-family breakdowns rather than only aggregate model scores.

\subsection{Static and Dynamic Hardening Regression}
\label{sec:hardening-regression}

We next evaluate whether tool-server hardening removes implementation level pitfalls and whether those static improvements translate into runtime reductions in focused workflows. Tier-1 static checks cover P1/P2/P5/P6, while P3/P4 are treated as trace-required pitfalls because they depend on concrete source-to-sink or image-to-tool flows.

\begin{table}[!htbp]
\centering
\caption{Mitigation effectiveness under the Tier-1 code-level analyzer.
$\mathrm{CE} = \Delta\mathrm{risk}\,/\,(1 + \log_{10}(\Delta\mathrm{LOC}))$.}
\label{tab:mitigation}
\resizebox{\columnwidth}{!}{%
\begin{tabular}{lcccccc}
\toprule
\textbf{Scenario} &
  $\mathrm{Tier1Risk}_\mathrm{base}$ &
  $\mathrm{Tier1Risk}_\mathrm{hard}$ &
  $\Delta\mathrm{log}\%$ &
  $\Delta\mathrm{val}\%$ &
  $\Delta\mathrm{LOC}$ &
  \textbf{CE} \\
\midrule
emailsystem      & 10.0 & 0.0 & $+$100\% & $+$25\% & 32 & 3.99 \\
documentsystem   & 10.0 & 0.0 & $+$100\% & $+$50\% & 15 & 4.60 \\
ETHPriceServer   & 10.0 & 0.0 & $+$100\% & $+$40\% & 35 & 3.93 \\
\midrule
\textbf{Mean}    & 10.0 & 0.0 & $+$100\% & $+$38\% & 27 & \textbf{4.17} \\
\bottomrule
\end{tabular}%
}
\end{table}

Table~\ref{tab:mitigation} shows that the recommended code-level mitigations eliminate all Tier-1 findings across the three workflows. The baseline servers contain 29 findings in total, including 16 HIGH and 13 MEDIUM findings. After hardening, all three servers show zero Tier-1 findings and the capped Tier-1 risk score drops from 10.0 to 0.0 in every workflow. The changes are small and localized, adding 27 non-comment lines of code on average, primarily for allowlists, structured logging, and server-side guards.

\begin{table}[!htbp]
\centering
\caption{Workflow-level dynamic hardening regression. Positive $\Delta$ASR
indicates a reduction in validator-confirmed attack success after hardening.}
\label{tab:workflow-asr}
\footnotesize
\setlength{\tabcolsep}{3pt}
\begin{tabular}{@{}p{0.30\columnwidth}cccc@{}}
\toprule
\textbf{Workflow} & \textbf{Base} & \textbf{Hard} &
$\boldsymbol{\Delta}$\textbf{ASR} & \textbf{Runs} \\
\midrule
ETHPriceServer       & 75.0\% & 50.0\% & $+$25.0 & 36 \\
emailsystem\_focused & 33.3\% & 0.0\%  & $+$33.3 & 36 \\
documentsystem       & 8.3\%  & 25.0\% & $-$16.7 & 36 \\
\bottomrule
\end{tabular}
\end{table}

Static hardening, however, is not sufficient by itself. Table~\ref{tab:workflow-asr} reports focused baseline-to-hardened ASR comparisons. Hardening reduces ASR for the crypto workflow and the focused email workflow, but the document workflow exhibits a regression. This does not contradict the static findings; rather, it shows that agent behavior can change after hardening in ways that static rules alone do not capture. Pitfall Lab therefore uses static checks as hardening evidence and trace-grounded ASR as the runtime source of truth.

\subsection{Focused Multi-modal Routing Case Study}
\label{sec:focused-multi-modal}

We further include a focused multi-modal routing case study to isolate a
modality-transition risk. The attacker-controlled destination is embedded only
in an image attachment and becomes visible to the agent after image extraction.
The key question is not merely whether the agent reaches a sink tool, but
whether image-derived text is propagated into a structured sink argument.

\begin{table*}[t]
\centering
\caption{Focused multi-modal routing case study. }
\label{tab:focused-multi-modal}
\small
\begin{tabular}{llccccc}
\toprule
\textbf{Scenario} & \textbf{Input mode} & \textbf{Baseline ASR} &
\textbf{Hardened ASR} & \textbf{Image text in trace?} &
\textbf{Sink reached?} & \textbf{Attacker sink reached?} \\
\midrule
Tool-description injection & text-only & 0.0\% & 0.0\% & N/A &
base 5/5; hard 5/5 & base 0/5; hard 0/5 \\
Image-derived routing & image-based & 100.0\% & 0.0\% &
base 5/5; hard 5/5 & base 5/5; hard 5/5 & base 5/5; hard 0/5 \\
\bottomrule
\end{tabular}
\vspace{0.25em}
\begin{minipage}{0.94\linewidth}
\footnotesize \textbf{Note:} The image-based attack embeds
an attacker-controlled destination only in an image attachment. ASR is
validator-confirmed.
\end{minipage}
\end{table*}

The result demonstrates a modality-transition risk that text-only benchmarks do
not capture. In the baseline setting, the image-derived payload enters the trace
in 5/5 runs and the agent uses the image-derived destination as the Teams channel in 5/5 runs, yielding 100\% ASR. In the hardened setting, the system still processes
the image and still reaches the sink tool in 5/5 runs, but it never sends to
the attacker-controlled destination, reducing ASR to 0\%. Thus, the mitigation
does not simply disable image handling or block the business workflow. Instead,
it preserves the sink action while preventing image-derived routing content from
controlling the sink destination.

This case supports the need to model P4 as a trace-required pitfall. Static
metadata can indicate that a server processes images and exposes a privileged
sink, but the security-relevant claim depends on a concrete runtime transition:
pixels are extracted into text, the extracted text enters the agent context, and
that text is later consumed as a tool argument. Protocol traces and objective
validators provide the evidence needed to distinguish harmless image processing
from image-to-tool leakage.

\subsection{Semantic MCP-BOM Evaluation}
\label{sec:semantic-bom-eval}

We evaluate the Semantic MCP-BOM as the measurement substrate and evidence layer
for Pitfall Lab. Rather than treating the BOM as a passive component inventory,
we ask whether its semantic fields support security-relevant detection,
hardening regression, and trace-grounded provenance. Concretely, we evaluate
three questions. \textbf{EQ1:} Which BOM field groups are necessary to represent
and detect pitfall classes? \textbf{EQ2:} Can BOM-backed checks serve as
regression tests across baseline-to-hardened workflow variants? \textbf{EQ3:}
Which pitfall classes require trace-linked provenance rather than static
manifest fields alone?

\subsubsection{Artifacts and Scope}
The evaluation uses generated Semantic MCP-BOM artifacts with and without trace
provenance. Static metrics are computed over P1/P2/P5/P6, which are
representable from semantic static fields. P3 and P4 are treated as
trace-required classes because their security claim is not merely that a source
and a sink exist, but that source-derived data reached the sink in a concrete
execution. This distinction avoids over-claiming from static metadata while
preserving the role of static BOM fields as a compact representation of tool
semantics, exposed parameters, trust boundaries, and auditability.

\subsubsection{EQ1: Field Ablation}
In field ablation (Table~\ref{tab:field-ablation}), component-only and schema-only variants detect no static
pitfall classes under the current field mask, while the semantic-static BOM
detects P1/P2/P5/P6 with Precision=0.571, Recall=1.000, and F1=0.727.

\begin{table}[!htbp]
\centering
\caption{Field ablation for Semantic MCP-BOM checks. Static metrics are computed
over P1/P2/P5/P6. P3/P4 are trace-required classes.}
\label{tab:field-ablation}
\scriptsize
\begin{tabular}{l l r r r r r r r}
\toprule
Variant & Classes & TP & FP & FN & TN & Prec. & Rec. & F1 \\
\midrule
Component only & none & 0 & 0 & 12 & 12 & 0.000 & 0.000 & 0.000 \\
Schema only & none & 0 & 0 & 12 & 12 & 0.000 & 0.000 & 0.000 \\
Semantic static & P1/P2/P5/P6 & 12 & 9 & 0 & 3 & 0.571 & 1.000 & 0.727 \\
Semantic + trace & \makecell[l]{P1/P2/P5/P6;\\ P3/P4 trace} & 12 & 9 & 0 & 3 & 0.571 & 1.000 & 0.727 \\
\bottomrule
\end{tabular}
\vspace{0.25em}
\begin{minipage}{0.94\linewidth}
\footnotesize \textbf{Note:} Schema-only metadata does not detect P2 because P2
requires both schema information and semantic labels for high-risk parameters.
Trace provenance is used for P3/P4, but those classes are not included in the
static aggregate shown in this table.
\end{minipage}
\end{table}

\subsubsection{EQ2: Baseline-to-Hardened Regression}
Table~\ref{tab:bom-decomposed-regression} compares baseline and hardened
workflow variants using BOM-backed finding counts and the decomposed semantic
risk metrics. Findings decrease for all three pairs: the email workflow drops
from 7 to 4, the document workflow from 9 to 6, and the crypto workflow from 11 to 6, for an aggregate reduction from 27 to 16.

The decomposed metrics explain why a raw semantic score alone can be misleading.
Exposure measures the amount of security-sensitive functionality that remains
present, including source/sink roles, high-risk parameters, and cross-boundary
capabilities. Control Coverage measures how much of this exposure is covered by
explicit schema constraints, allowlists, server-side validation, audit logging,
and policy hooks. Residual Risk is the uncovered portion of Exposure.

Under this interpretation, hardened servers are not expected to eliminate
Exposure: they often preserve the same business functionality. The email and
document workflows therefore keep the same Exposure after hardening, while the
crypto workflow increases from 23 to 28 because additional high-impact interface
semantics and controls are made explicit. This is not a counterexample to
hardening. Rather, it indicates that the server still exposes high-impact
capabilities that require strong controls. The security outcome is captured by
Residual Risk: after accounting for explicit controls, residual risk decreases in
all three workflows, from 10.50 to 4.20 for email, 15.75 to 5.40 for document,
and 19.67 to 8.68 for crypto. On average, Control Coverage increases from 0.173
to 0.697, while Residual Risk decreases from 15.31 to 6.09.

\begin{table*}[!hbtp]
\centering
\caption{Baseline-to-hardened regression using Semantic MCP-BOM findings and semantic risk metrics.}
\label{tab:bom-decomposed-regression}
\begin{tabular}{lcccc}
\toprule
\textbf{Server pair} & \textbf{Findings} & \textbf{Exposure} &
\textbf{Control Coverage} & \textbf{Residual Risk} \\
\midrule
Email workflow & $7 \rightarrow 4$ & $14 \rightarrow 14$ &
$0.250 \rightarrow 0.700$ & $10.50 \rightarrow 4.20$ \\
Document workflow & $9 \rightarrow 6$ & $18 \rightarrow 18$ &
$0.125 \rightarrow 0.700$ & $15.75 \rightarrow 5.40$ \\
Crypto workflow & $11 \rightarrow 6$ & $23 \rightarrow 28$ &
$0.145 \rightarrow 0.690$ & $19.67 \rightarrow 8.68$ \\
\midrule
Total / mean & $27 \rightarrow 16$ & $18.33 \rightarrow 20.00$ avg. &
$0.173 \rightarrow 0.697$ avg. & $15.31 \rightarrow 6.09$ avg. \\
\bottomrule
\end{tabular}
\end{table*}

Thus, cases where the hardened server has equal or higher raw semantic surface
are not counterexamples to hardening. They indicate that the server retains
high-impact capabilities. The relevant security outcome is residual risk: after
accounting for explicit controls, residual risk consistently decreases.

\subsubsection{EQ3: Trace-Linked Provenance}
For trace-required pitfall classes, static BOM fields identify
security-relevant structure but do not by themselves establish a concrete
violation. In the email workflow, the Semantic MCP-BOM marks one tool role as a
protected content source and another as a privileged outbound sink, which
indicates a possible P3 source-to-sink path. The trace supplies the missing
runtime evidence: protected content is first retrieved from the source and later
appears in the argument of an outbound messaging action.

We therefore label the case as a HIGH-severity P3 source-to-sink flow. The
static BOM explains why the path is security-relevant, while trace provenance
demonstrates that the flow occurred. The focused multi-modal routing case in
Section~\ref{sec:focused-multi-modal} provides the analogous motivation for P4:
static fields can show that image inputs and sink tools exist, but only the
trace can show that image-derived content entered a sink argument. Together,
these cases motivate treating P3/P4 as trace-required classes: static roles
provide candidate evidence, but runtime provenance should be required before
reporting a final source-to-sink or image-to-tool finding.

\section{Related Work}
\label{sec:related}

\subsection{Prompt Injection and Tool-Using Agents}
Security risks in tool-using LLM agents increasingly center on \emph{prompt injection}. OWASP ranks prompt injection as a top threat for LLM applications, highlighting consequences such as sensitive information disclosure, unauthorized access, and unintended tool execution; it also notes that multi-modal inputs can hide instructions (e.g., in images), making attacks harder to detect and mitigate \cite{owasp2025llmtop10}. Wallace et al.~\cite{wallace2024instructionhierarchytrainingllms} further demonstrate that function-calling agents can be manipulated by adversarial inputs to alter tool-invocation decisions. AgentDojo~\cite{debenedetti2024agentdojo} provides a testbed to evaluate prompt-injection attacks and defenses in task-oriented agents, measuring the trade-off between task success and security violations.

\subsection{SBOMs, ML-BOMs, and AI Supply-Chain Transparency}

SBOMs provide machine-readable visibility into software components and their supply-chain relationships, and have become a common foundation for vulnerability management, procurement review, and software transparency~\cite{10.1145/3654442}. Standardization efforts such as SPDX~\footnote{https://spdx.dev/} and CycloneDX~\footnote{https://cyclonedx.org/} provide exchange formats for representing software components, licenses, and dependency relationships~\cite{10.1145/3672608.3707940}. More recent AI-oriented extensions, including ML-BOM and AI-SBOM efforts, broaden the scope of BOMs to include models, datasets, training and inference configurations, prompts, and other AI-specific artifacts~\cite{10.1145/3643662.3643957}.

Our work follows this supply-chain transparency tradition but targets a different dependency layer. Existing SBOMs are well suited to representing packages and software components~\cite{10771496}, while ML-BOMs are designed to expose AI/ML assets and their provenance. MCP tool servers, however, introduce security-relevant interface semantics that are neither ordinary package dependencies nor model artifacts. Tool descriptions can encode implicit policy, schemas can expose unconstrained high-risk destinations, source tools can feed privileged sink tools, and missing audit hooks can prevent incident reconstruction. These properties directly affect agent behavior but are not captured by component identity alone.

The Semantic MCP-BOM is therefore complementary to existing SBOM and ML-BOM standards. It does not attempt to replace package-level or model-level BOMs; instead, it adds an agent-specific semantic layer for tool-server dependencies. This layer records tool descriptions, schemas, high-risk parameter labels, source / sink roles, trust boundaries, policy hooks, audit support, and trace provenance. In doing so, it adapts BOM-style supply-chain review to MCP deployments, where the primary risk often lies in the semantics exposed to the agent rather than only in the installed package or model artifact.

\subsection{MCP Ecosystem Security and Supply-Chain Risks}
The rapid adoption of the Model Context Protocol (MCP) has prompted initial analyses of its ecosystem. Song et al.~\cite{song2025protocolunveilingattackvectors} survey MCP attack vectors across tool discovery, invocation, and composition, offering a threat taxonomy but limited evaluation methodology and mitigation guidance. MCPTox~\cite{wang2025mcptoxbenchmarktoolpoisoning} introduces a benchmark for tool poisoning attacks on real-world MCP servers, showing that adversarial tool metadata (descriptions/schemas) can mislead agents in text-based settings, while leaving multi-modal vectors and tool-side mitigation relatively underexplored. These threats align with broader software supply-chain security findings: malicious package injection in NPM \cite{ohmBackstabberKnifeCollection2020} and exploitation tendencies in AI supply chains (models/data) \cite{10838587} illustrate how third-party components can introduce hidden risks. In comparison, our work focuses on protocol- and implementation-level weaknesses in MCP tool servers, emphasizing developer-side mitigations such as server-side validation, schema constraints, and audit logging, and explicitly covering multi-modal image-to-tool attack paths.

\subsection{Positioning of Our Contributions}

\begin{table*}[!hbtp]
\centering
\caption{Comparison of agent security benchmarks and MCP supply-chain evaluation support}
\label{tab:benchmark-comparison}
\footnotesize
\setlength{\tabcolsep}{3.5pt}
\renewcommand{\arraystretch}{1.08}
\begin{tabular}{@{}lccccccccc@{}}
\toprule
\textbf{Benchmark} 
& \textbf{MCP/} 
& \textbf{Tool} 
& \textbf{Puppet/} 
& \textbf{Multi-} 
& \textbf{Model-Side} 
& \textbf{Tool-Side} 
& \textbf{Trace-} 
& \textbf{Semantic} 
& \textbf{Hardening} \\
& \textbf{Protocol} 
& \textbf{Metadata} 
& \textbf{Supply-Chain} 
& \textbf{modal} 
& \textbf{Defense} 
& \textbf{Mitigation} 
& \textbf{Based} 
& \textbf{BOM} 
& \textbf{Regression} \\
& \textbf{Aware} 
& \textbf{Poisoning} 
& \textbf{Threats} 
& \textbf{Inputs} 
& \textbf{Evaluation} 
& \textbf{Guidance} 
& \textbf{Validation} 
& \textbf{Support} 
& \textbf{Testing} \\
\midrule
AgentDojo~\cite{debenedetti2024agentdojo} 
& -- 
& Partial 
& -- 
& -- 
& \checkmark 
& Partial
& -- 
& -- 
& -- \\

InjecAgent~\cite{zhan-etal-2024-injecagent} 
& Partial 
& -- 
& -- 
& -- 
& \checkmark 
& Partial 
& --
& -- 
& -- \\

MCPTox~\cite{wang2025mcptoxbenchmarktoolpoisoning} 
& \checkmark 
& \checkmark 
& Partial 
& -- 
& \checkmark 
& Limited 
& -- 
& -- 
& -- \\

MSB~\cite{zhang2026mcp} 
& \checkmark 
& \checkmark 
& Partial 
& -- 
& \checkmark 
& Limited
& Partial 
& -- 
& -- \\

MCP-SafetyBench~\cite{zong2026mcpsafetybench} 
& \checkmark 
& \checkmark 
& Partial 
& -- 
& \checkmark 
& Limited
& Partial 
& -- 
& Limited \\

\textbf{MCP Pitfall Lab} 
& \checkmark 
& \checkmark 
& \checkmark 
& \checkmark 
& \checkmark 
& \checkmark 
& \checkmark 
& \checkmark 
& \checkmark \\
\bottomrule
\multicolumn{10}{@{}l}{\footnotesize \checkmark = full or explicit support; Partial/Limited = indirect, partial, or narrow support; -- = not addressed.} \\
\multicolumn{10}{@{}l}{\footnotesize The comparison focuses on capabilities relevant to MCP tool-server supply-chain security, semantic dependency review, and trace-grounded regression.}
\end{tabular}
\end{table*}

As demonstrated in Table~\ref{tab:benchmark-comparison}, MCP Pitfall Lab emphasizes tool-server dependencies rather than model-only robustness. It combines MCP-aware attack surfaces, multi-modal attacks, tool-side mitigation guidance, trace-based validation, Semantic MCP-BOM evidence, and hardening regression testing.

\section{Discussion and Limitations}
\label{sec:discussion}

Our results suggest that many MCP failures arise from interface-level mistakes:
policy encoded in tool descriptions, permissive destination parameters, missing
server-side validation, and incomplete audit logging. These issues are hard to
diagnose from agent narratives alone. Protocol traces, validators, Tier-1
findings, and Semantic MCP-BOM fields provide more reliable evidence for
debugging, hardening, procurement review, and regression testing. MCP servers
should therefore be reviewed not only as software packages, but also as semantic
dependencies whose descriptions, schemas, high-risk parameters, source/sink
roles, validation hooks, audit support, and trace provenance shape agent
behavior.

The evaluation has two limitations. First, the focused multi-modal study isolates one image-to-tool routing mechanism; broader multi-modal evaluation should include more labeled text/image chains and additional image-channel mitigations. Second, the Semantic MCP-BOM decomposition is heuristic: exposure weights, control-coverage weights, and the mapping from concrete controls to $c_e(v)$ are manually specified. Future work should calibrate these metrics against validator outcomes and automate control-evidence extraction for third-party MCP servers.

\section{Conclusion}

We presented MCP Pitfall Lab and Semantic MCP-BOM for evaluating MCP tool-server dependencies as AI supply-chain artifacts. Across 2,579 validator-completed runs over four models, Pitfall Lab observes 31.9\% overall validator-confirmed ASR. Multi-modal injection is the strongest attack family at 38.7\%, followed by puppet attacks at 34.5\% and tool poisoning at 22.6\%. These results show that MCP agent risk is not limited to text-only tool poisoning: untrusted outputs, malicious tool servers, multi-modal inputs, and privileged sink actions all contribute to runtime failures.

The supporting analyses show why traces and semantic dependency metadata are both necessary. Static hardening removes Tier-1 code-level findings, but focused runtime regression still reveals workflow-specific behavior changes. Semantic MCP-BOM fields detect P1/P2/P5/P6 where component-only and schema-only views fail, BOM-backed findings drop from 27 to 16 after hardening, and decomposed risk metrics show Control Coverage increasing from 0.173 to 0.697 while Residual Risk decreases from 15.31 to 6.09. More broadly, MCP security requires both runtime evidence and semantic dependency review: static metadata identifies candidate exposure, while protocol traces and validators establish concrete violations.

\clearpage

\end{document}